# Polytropic Behavior of Solar Wind Protons Observed by *Parker Solar Probe*

Georgios Nicolaou[1*], George Livadiotis[2], Robert T. Wicks[1], Daniel Verscharen[1,3], Bennett A. Maruca[4]

[1]Department of Space and Climate Physics, Mullard Space Science Laboratory, University College London, Dorking, Surrey, RH5 6NT, UK

[2]Division of Space Science and Engineering, Southwest Research Institute, San Antonio, TX 78238, USA

[3]Space Science Center and Department of Physics, University of New Hampshire, Durham, NH 03824, USA

[4]Bartol Research Institute, Department of Physics and Astronomy, University of Delaware, Newark, DE 19716, USA

**Abstract**

A polytropic process describes the transition of a fluid from one state to another through a specific relationship between the fluid density and temperature. The value of the polytropic index that governs this relationship determines the heat transfer and the effective degrees of freedom during the process. In this study, we analyze solar wind proton plasma measurements, obtained by the Faraday cup instrument on-board *Parker Solar Probe*. We examine the large-scale variations of the proton plasma density and temperature within the inner heliosphere explored by the spacecraft. We also address a polytropic behavior in the density and temperature fluctuations in short-time intervals, which we analyze in order to derive the effective polytropic index of small time-scale processes. The large-scale variations of the solar wind proton density and temperature which are associated with the plasma expansion through the heliosphere, follow a polytropic model with a polytropic index ~5/3. On the other hand, the short time-scale fluctuations which may be associated with turbulence, follow a model with a larger polytropic index. We investigate possible correlations between the polytropic index of short time-scale fluctuations and the plasma speed, plasma beta, and the magnetic field direction. We discuss the scenario of mechanisms including energy transfer or mechanisms that restrict the particle effective degrees of freedom.

*Corresponding author g.nicolaou@ucl.ac.uk

## 1. Introduction

Polytropic process is any thermodynamic process in which the density *n* and temperature *T* (or pressure *P*) of a fluid are related through the polytropic index *γ* such as

$$T \propto n^{\gamma-1} \text{ or } P \propto n^{\gamma}. \qquad (1)$$

During a polytropic process, the ratio of the energy transferred in the system as heat over the energy transferred as work remains constant (e.g. Parker 1963; Chandrasekhar 1967). In the special case where there is no heat transfer in the system during the transition, the process is called adiabatic. In an adiabatic process, *γ* is equal to the ratio of the specific heats $c_p/c_v$ and is related with the degrees of freedom *f* of the plasma particles, such as $\gamma = c_p/c_v = 1 + \frac{2}{f}$. The knowledge of *γ* is essential for the fluid (including MHD) description of plasma processes and thus, necessary in understanding several physical processes in plasma environments. The polytropic equation brings closure to the moments hierarchy by connecting higher order moments (*T*, *P*) with the $0^{th}$ order moment (*n*) of the velocity distribution function of plasma particles (e.g., Kuhn et al. 2010). Moreover, through the polytropic equation we understand physics of transitions in the medium without the need to solve the explicit energy equation describing the system. For instance, *γ* defines the compression ratio of shocks (e.g., Parker 1961; Livadiotis 2015; Scherer et al. 2016; Nicolaou & Livadiotis 2017), the nature of turbulent fluctuations (e.g., Bavassano et al. 1996, Verscharen et al. 2016, 2017, 2019; Wu et al. 2019), the expansion of plasma within several environments, such us planetary magnetospheres (e.g., Spreiter & Stahara 1994), the interplanetary space (e.g., Elliott et al. 2019), the heliosheath (e.g., Livadiotis et al. 2011; Livadiotis & McComas 2013), and more. Importantly, recent studies argue that the fluid approach describes successfully small time-scale fluctuations, even in plasmas with low collisionality, such as the plasma protons of solar wind and the terrestrial magnetosheath (e.g., Verscharen et al. 2017, 2019; Wu et al. 2019). Therefore, the polytropic description is possibly applicable within a wide range of plasma scales.

The value of *γ* characterizes specific processes in individual fluid parcels and thus, it may be different within different plasma regimes, different plasma species, and/or it may vary with time. Moreover, Livadiotis 2016 analyze streamlines consist of multiple polytropes. Some studies investigate the polytropic behavior of plasmas in large scale processes, such as the solar wind expansion within the heliosphere. For instance, Totten et al. 1995 determine *γ* through the radial profiles of the solar wind proton density and temperature within a wide range of heliocentric distances. The derived *γ* lead to the conclusion that the solar wind plasma protons are heated as they propagate in the outer heliosphere.

In other typical analyses, *γ* is determined in individual streamlines where Eq.(1) is valid. For example, studies determine the polytropic behavior in streamlines of space plasma species within several regimes, such us planetary magnetospheres (e.g., Arridge et al. 2009; Nicolaou et al. 2014a; Dialynas et al. 2018; Pang et al. 2015; Park et al. 2019), magnetic clouds (e.g., Osherovich et al. 1993), the solar wind (e.g., Newbury et al. 1997; Kartalev et al. 2006; Nicolaou et al. 2014b, 2019; Livadiotis & Desai 2016; Livadiotis 2018a, 2018b; Nicolaou & Livadiotis 2019; Elliott et al. 2019), and the inner heliosheath (e.g., Livadiotis et al. 2013).

In this study, we analyze observations by NASA's *Parker Solar Probe* (*PSP*). We use data obtained within ~0.17 au and ~ 0.8 au, giving the opportunity to investigate the macroscopic changes of the plasma parameters as the solar wind expands into the heliosphere. Moreover, the high-time-resolution measurements of solar wind protons, allow the investigation of the polytropic indices in very small time-scales, associated with plasma compressive fluctuations due to waves and turbulence. We also investigate if the polytropic behavior of the plasma protons depends on the solar wind speed $U$, which is often used as the basic criterion to distinct between solar wind plasmas of different solar origins (e.g., McComas et al. 2003) with different properties (e.g., Geiss et al. 1995; Marsch et al. 1982, 1983; Hellinger et al. 2011; Borovsky 2016; Perrone et al. 2019; Stansby et al. 2019; Huang et al. 2020). Finally, we examine the polytropic index as a function of the plasma beta, and the magnetic field direction. In the next section we describe the data-set we use in this study. In Section 3, we describe our methods and Section 4 shows our results, which we discuss in Section 5. Finally, in Section 6, we summarize our conclusions.

## 2. Data

*Solar Probe Cup* (*SPC*) of the Solar Wind Electrons Alphas and Protons (SWEAP) suite (Kasper et al. 2016), is a Sun-pointing Faraday cup capturing the mainly radial bulk flow of the plasma particles. The integrated over the field of view (~60°) flux is converted to 1D distribution and analyzed to derive the plasma bulk properties. Here, we use the plasma proton density, temperature, and speed as derived from fittings to the spectra observed by *SPC*. We use the plasma observations obtained between 2018 November 6 and 2019 May 16. The specific time interval includes two approaches to the Sun at ~ 0.17 au. We separate the selected data-set in three consecutive time intervals, in a way that each interval covers the maximum possible heliocentric distance range (see bottom panel of Figure 1). Large ranges in heliocentric distance correspond to large variations in the plasma density and temperature allowing the investigation of the polytropic relation within the inner heliosphere. During each interval, the spacecraft samples streamlines over a wide range of Carrington longitude (top panel of Figure 1). Interval 1 starts at the perihelion on 2018 November 6 and ends at the aphelion on 2019 January 20, when the second interval begins. Interval 2 ends at the perihelion on 2019 April 4, when the third interval begins. Interval 3 ends on 2019 May 15. We use only the plasma measurements which are flagged as "measurements with no conditions" and with relative density and temperature uncertainty $\sigma_n/n$ and $\sigma_T/T$ < 30% respectively. The specific selection criterion is met in ~29% of the data obtained within the time period we examine. For our investigation of the polytropic behavior dependence on the plasma beta and the magnetic field direction which may affect the derived temperature (see Section 4.6), we use a 2-day time interval, between 2019 April 2 and 2019 April 4. For the specific case study, we use high-time-resolution magnetic field observations by the Electromagnetic Fields Investigation (*FIELDS*, Bale et al. 2016), obtained within the specific interval.

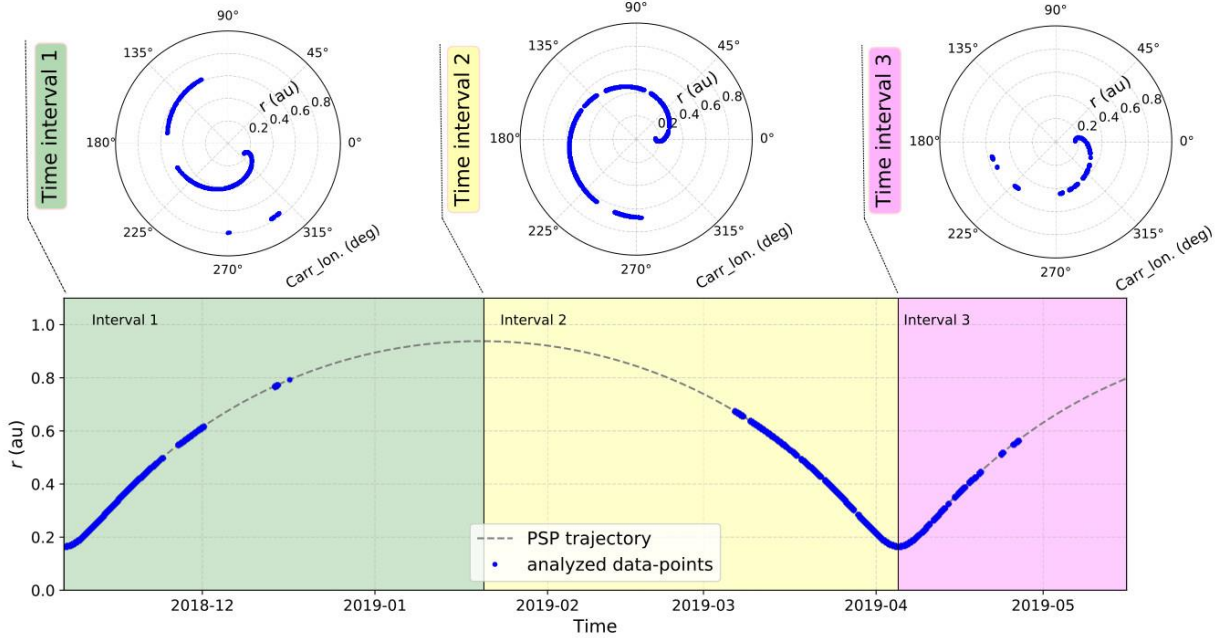

**Figure 1**. The heliocentric distance of *PSP* from 2018 November 6 to 2019 May 15. The shadowed regions correspond to the three time intervals we analyze separately in this study. The blue data-points correspond to the observations with no warning flags and small relative density and temperature uncertainties ($\sigma_n/n$ and $\sigma_T/T < 0.3$), which we analyze in this study. The panels at the top show the spacecraft longitude in the Carrington coordinate system for the three time intervals.

### 3. Methodology

According to Eq.(1):

$$\log_{10}(T) = (\gamma - 1)\log_{10}(n) + C, \qquad (2)$$

where the constant *C* is generally different for different plasma streamlines. We initially examine the two-dimensional (2D) histograms of $\log_{10}(n)$ and $\log_{10}(T)$ in order to investigate if the large-scale plasma expansion in the heliosphere follows the polytropic model in Eq.(2). As different streamlines are crossing the spacecraft during the examined intervals, the linear relationship in Eq.(2) can be determined only if the change of *C* across the different streamlines is smaller than the large-scale variations of $\log_{10}(n)$ and

log$_{10}$(T). Therefore, we do not attempt to determine a unique polytropic relation using the large-scale variations of the plasma. Instead, we calculate $\gamma$ and its one-sigma error $\sigma_\gamma$ in small timescales, by dividing each interval to smaller consecutive sub-intervals and performing a weighted orthogonal-distance regression to $\ln T$ vs $\ln n$ within each sub-interval. We use weighted orthogonal regression, considering symmetric uncertainty $\sigma_{\ln T} \sim \sigma_T/T$ and $\sigma_{\ln n} \sim \sigma_n/n$ for each $\ln T$ and $\ln n$ data-point respectively. We specifically examine sub-intervals of 8s, 64s, and 512s. The plasma fluctuations in such a small time scales are mainly due to transient compressions and compressive fluctuations associated with wave-particle interactions. To increase the statistical significance of the results, we reject sub-intervals with less than five data points and those for which $\sigma_\gamma > 1$. The short length of the examined sub-intervals eliminates the possibility to mix observations of different streamlines (e.g., Kartalev et al. 2006; Pang et al. 2015; Nicolaou et al. 2014b, 2019; Nicolaou & Livadiotis 2019). In Figure 2, we show an example of $\ln T$ as a function of $\ln n$ within a typical 8s sub-interval (left) and a typical 64s sub-interval (right). The data-points within the shown subintervals have a clear linear behavior described by the fitted model in Eq.(2) to the observations. The slope of the fitted line in each subinterval determines $\gamma$. The fitted lines to the sub-intervals in Figure 2, have the same slope (~1.7), which corresponds to $\gamma \sim 2.7$. The two lines however, have a different y-intersect, corresponding to a different $C$ constant in each sub-interval.

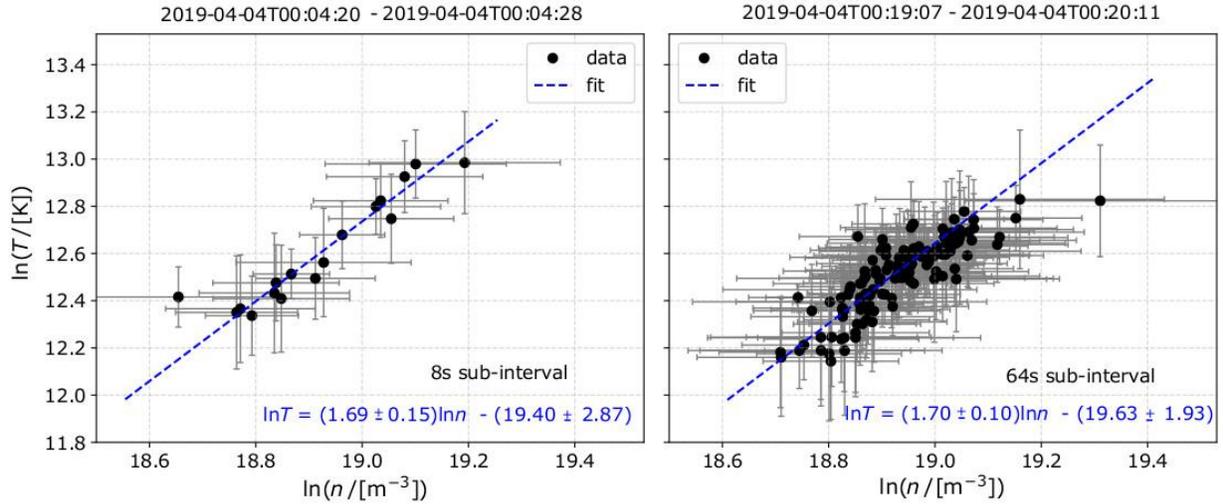

**Figure 2**. $\ln T$ as a function of $\ln n$ within a typical sub-interval of (left) 8 seconds and (right) 64 seconds. The vertical (horizontal) error bars correspond to $\sigma_T/T$ ($\sigma_n/n$). In both panels, the blue line is the linear fitting of Eq.(2) to the data-points. The slope of the fitted line determines $\gamma$. In both examples, the slope of the fitted line is ~1.7 which corresponds to $\gamma \sim 2.7$.

## 4. Results

### *4.1 Histograms of Large-Scale Variations*

In Figure 3, we show the 2D histograms of $\log_{10}(n)$ and $\log_{10}(T)$ for the three time intervals shown in Figure 1. In each panel we draw adiabatic models for protons with three degrees of freedom ($f = 3$, $\gamma = 5/3$) as guides to the eye. As mentioned in Section **3**, different plasma streamlines pass the spacecraft within the sampled heliocentric distance range and it is impossible to determine a single polytropic relationship describing the whole data-set within each interval. On the other hand, we investigate if the statistical sampling of streamlines within the examined heliocentric distance range can reveal a characteristic slope, and hence a polytropic behavior that describes the large scale variations of $\log_{10}(n)$ and $\log_{10}(T)$ associated with the expanding solar wind in the heliosphere. Within each interval we examine here, the proton density ranges roughly between ~10 cm$^{-3}$ and ~1000 cm$^{-3}$, while the proton temperature ranges between ~$3\times10^4$ K and ~$3\times10^6$ K. The large-scale variations of *n* and *T* within all the three intervals tend to follow the adiabatic model, especially for $n < 100$ cm$^{-3}$. Within the higher density regime ($n > 100$ cm$^{-3}$) the temperature seems to drop below the adiabatic model, especially in interval 1 and 3, while in interval 2 we still observe high occurrence along the adiabatic line. Moreover, within intervals 1 and 2, we observe discrete "bright" structures that deviate from the $\gamma = 5/3$ behavior as their slopes are quite larger than the slope of the adiabatic model.

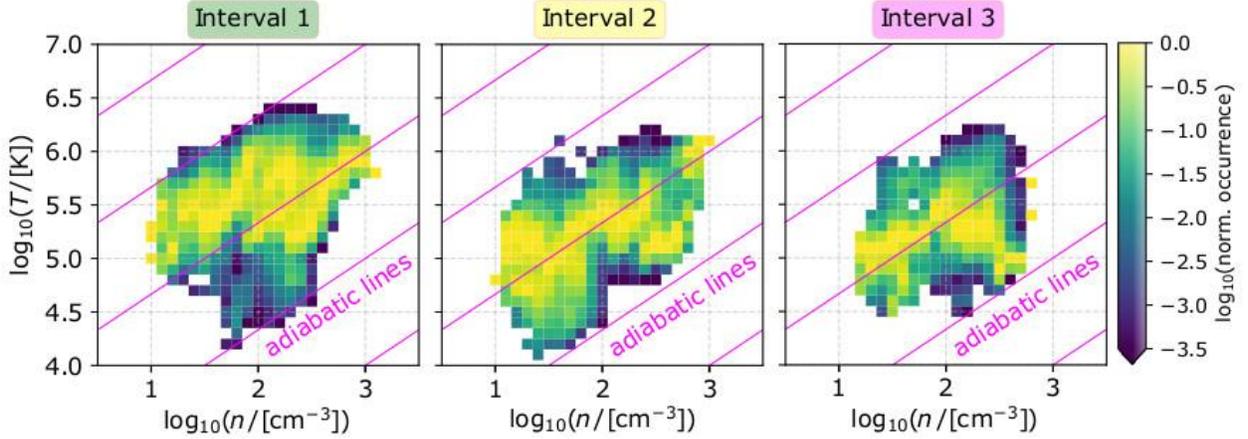

**Figure 3**. Two-dimensional histograms of $\log_{10}(n)$ and $\log_{10}(T)$ for the three time intervals shown in Figure 1. In each panel we show the adiabatic lines for protons with three degrees of freedom ($f = 3$, $\gamma = 5/3$). Typically, the large-scale changes of the plasma density and temperature within each sub-interval follow a near-adiabatic trend. However, we observe features with $\gamma > 5/3$.

*4.2 Density and Temperature vs radial distance*

As the solar wind expands in the heliosphere the density decreases. For a polytropic expansion, the plasma temperature will change accordingly. In the top panel of Figure 4, we show the 2D histogram of the proton number density and radial distance for interval 1. The magenta line shows the model for uniform expansion with constant radial velocity $n \propto r^{-2}$. Although $n \propto r^{-2}$ captures basic features of the entire profile, we can distinct regions where the decrement of $n$ with $r$ is either steeper or more gradual. The bottom panel of Figure 4 shows the 2D histogram of temperature and the radial distance for the same interval. The magenta line shows an expansion model for $\gamma = 5/3$, assuming a constant radial speed, such as $n \propto r^{-2}$. The grey line shows a radial profile with same slope as the one characterizing the parallel proton temperature of fast solar wind protons, determined by Huang et al. 2020 from the analysis of *SPC* data. Both of these models seem to capture the large-scale features of the entire profile. However, there are distinct sub-intervals with steeper slopes which are better described with the blue lines showing radial expansion models with $\gamma = 2.7$, which as we show in the next section, is a typical value within short time sub-intervals of the analyzed data-set. We discuss further our results and their possible implications in our discussion section.

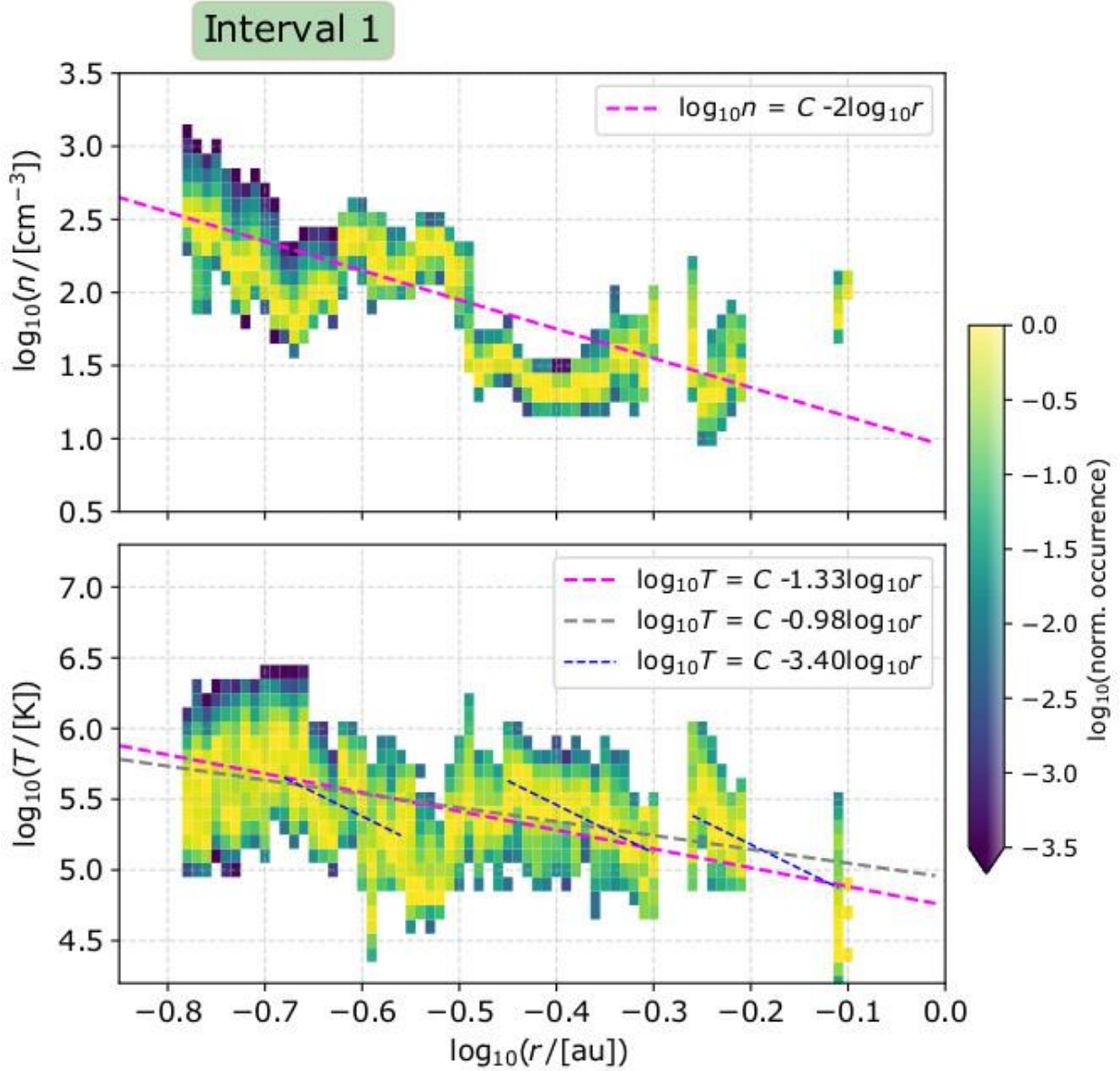

**Figure 4.** Two-dimensional histograms of (top) the proton density and (bottom) the proton temperature as functions of the radial distance for time interval 1. The magenta line on the top panel shows the expected density for an expansion model with constant speed $n \propto r^{-2}$. In the lower panel, the magenta line shows the expected temperature of a polytropic radial expansion model with $\gamma = 5/3$ while the blue lines are expansion models for $\gamma = 2.7$. The grey line is a function with the slope determined in the analysis by Huang et al. 2020 for the parallel proton temperature of fast solar wind observed by *SPC*.

*4.3 Small-Scale Regression*

As the large scale variations of *n* and *T* are associated with the plasma expansion within the heliosphere, short time-scale fluctuations are associated with other physical processes, such as wave-particle interactions. Thus, for a conclusive study of the polytropic behavior of solar wind protons, we analyze short time intervals. The selection of short time intervals also eliminates the spacecraft crossings between multiple streamlines. We manage to analyze sub-intervals with a clear polytropic behavior by applying the selection criteria explained in Section 3. In Figure 5, we show the histograms of *γ*, as derived in sub-intervals of 8s, 64s, and 512s within each of the time intervals 1, 2, and 3. In Table 1, we show the most frequent values and standard deviations of the histograms in each interval. The most frequent value is *γ* ~ 2.7 for the majority of the examined histograms, while the standard deviation progressively increases with increasing length of the sub-intervals. Within interval 1, the standard deviation of the polytropic index is $\sigma_\gamma$ ~ 0.80 when determined in 8s sub-intervals, $\sigma_\gamma$ ~ 1.05 in 64s sub-intervals, and $\sigma_\gamma$ ~ 1.80 in 512s sub-intervals. Within interval 2, $\sigma_\gamma$ ~ 0.67 in 8s sub-intervals and progressively increases to $\sigma_\gamma$ ~ 1.85 in 512s sub-intervals. Finally, within interval 3, the analysis of 8s sub-intervals derives $\sigma_\gamma$ ~ 0.64, which increases progressively to $\sigma_\gamma$ ~ 1.90 in 512s sub-intervals.

As we already mention, by increasing the subinterval length we increase the chance of streamline mixing. The linear fitting of Eq.(2) to ln*T* - ln*n* data-points of different streamlines, fails to derive accurately the index *γ*, if the mixed streamlines have either different *γ*, or/and a different constant *C*. Such artifacts are possibly responsible for the asymmetry of the histograms in Figure 5. Since the possibility of streamline mixing is reduced for the shortest sub-interval length, hereafter, we use the 8s sub-interval analysis results to explore further the polytropic behavior of the solar wind protons.

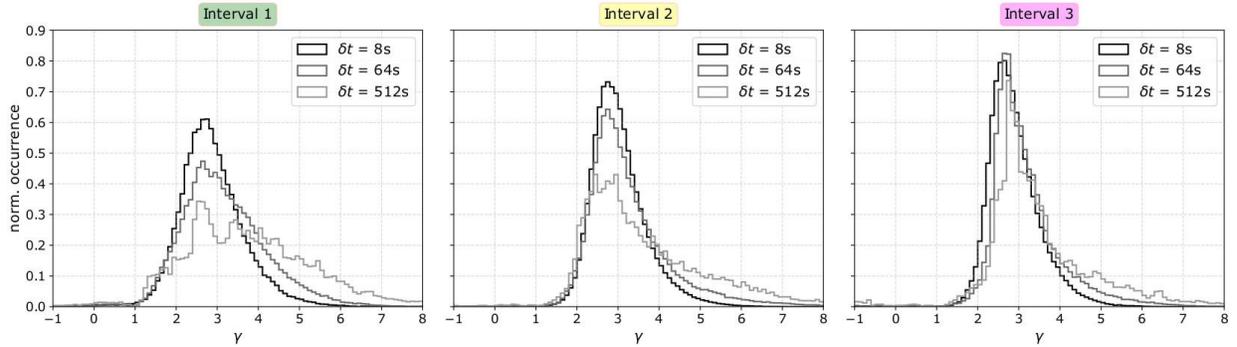

**Figure 5**. Histograms of *γ* as derived from the linear regression of short time-scale sub-intervals (8s, 64s, and 512s) within each of the three intervals shown in Figure 1, and 3. The most frequent value is *γ* ~ 2.7, for every timescale we examine and $\sigma_\gamma$ increases with increasing length *δt* of the analyzed sub-intervals.

**Table 1**

The most frequent values and standard deviations of the $\gamma$ histograms shown in Figure 4.

| Interval name | Sub-interval duration $\delta t$ (s) | Most frequent $\gamma$ value | Standard deviation $\sigma_\gamma$ |
|---|---|---|---|
| Interval 1 | 8 | 2.7 | 0.8 |
|  | 64 | 2.7 | 1.1 |
|  | 512 | 2.6 | 1.8 |
| Interval 2 | 8 | 2.7 | 0.7 |
|  | 64 | 2.7 | 1.1 |
|  | 512 | 2.5 | 1.9 |
| Interval 3 | 8 | 2.7 | 0.6 |
|  | 64 | 2.7 | 1.0 |
|  | 512 | 2.8 | 1.9 |

*4.4 Polytropic Behavior vs Speed*

We examine the polytropic behavior of the solar wind as a function of its speed, in an attempt to investigate if plasmas of different origins follow different polytropic processes. In the left panel of Figure 6, we show the 2D histogram of $\gamma$ as derived from the analysis of 8s sub-intervals within interval 1, and the average $U$ within the corresponding sub-intervals. On the right and top side of the panel, we show the 1D histograms of $\gamma$ and $U$, respectively. The right panel of Figure 6 shows the 2D histogram of $\gamma$ and $U$, normalized to its maximum occurrence in each $U$ bin. The magenta line in both panels shows the most frequent value of $\gamma$ as a function of $U$. The speed within this interval ranges between ~ 250 kms$^{-1}$ and ~700 kms$^{-1}$. The most frequent $\gamma$ ~ 2.7 does not exhibit any systematic variations with speed.

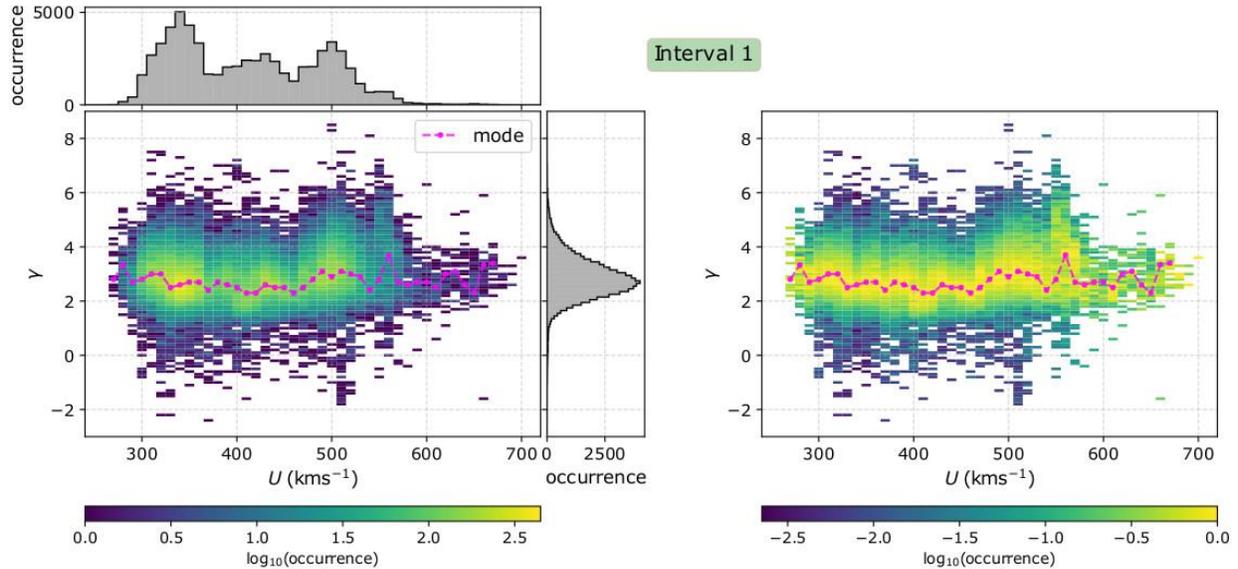

**Figure 6**. (Left) Two-dimensional histogram of *γ* and *U* within interval 1. On the right and top of the panel we show the 1D histograms of *γ* and *U* respectively. (Right) the 2D histogram of *γ* and *U*, normalized to the maximum value per *U* bin. The magenta line in both panels shows the most frequent value of *γ* in each *U* bin.

### *4.5 Polytropic Behavior vs Plasma Beta*

The proton plasma beta $\beta = 2\mu_0 nkT/B^2$, shows how energy densities are partitioned between the plasma pressure ($P = nk_BT$, with $k_B$ the Boltzmann constant) and the magnetic field pressure ($P_B = B^2/(2\mu_0)$, with $\mu_0$ the vacuum permeability). Since we investigate thermodynamic processes in a magnetized plasma it is important to explore the role of the plasma beta, which we know it plays a vital role in wave-particle interactions (e.g. Bruno & Carbone 2013; Verscharen et al. 2019, and references therein. In the left panel of Figure 7, we show the 2D histogram of *γ* and $\log_{10}\beta$ derived from the analysis of 8s sub-intervals between 2019 April 2 and 2019 April 4. For the specific case study, we calculate *β* from the average *n*, *T*, and *B* over each 8s sub-interval to much the time resolution of the derived *γ*. On the right and top side of the 2D histogram, we show the 1D histogram of *γ* and the 1D histogram of $\log_{10}(\beta)$ for the same time period, respectively. The right panel of Figure 7, shows the 2D histogram of *γ* and $\log_{10}\beta$, normalized to the maximum value in each $\log_{10}(\beta)$ bin. In both panels, the magenta line shows the most frequent value of *γ* as a function of $\log_{10}(\beta)$. Within the analyzed time period, *β* ranges between ~0.08 and ~10. We distinct two regions in the 2D histograms where *γ* decreases with *β*. More specifically, *γ* decreases from ~3.2 to ~2.5 as *β* increases from 0.08 to 0.3. At *β* ~ 0.3, there is a sharp increase of *γ* from 2.5 to ~3, which may be associated with a crossing between different plasma parcels (streamlines). Finally, *γ* decreases from 3 to ~1.8 as *β* increases from ~0.4 to 10.

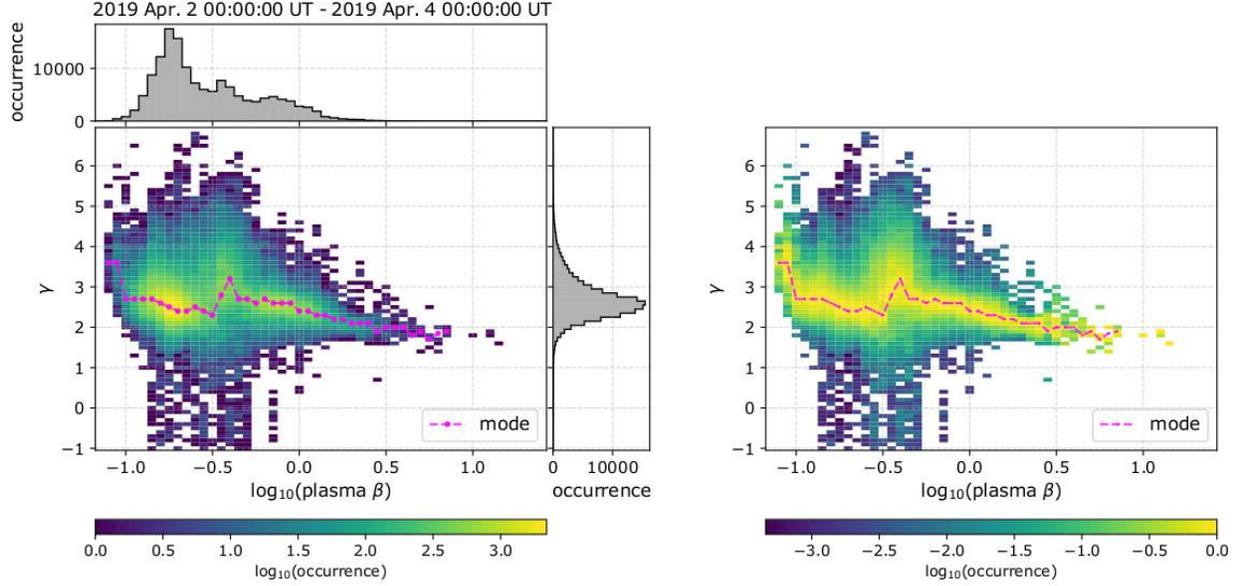

**Figure 7**. (Left) Two-dimensional histogram of $\gamma$ and $\log_{10}(\beta)$ for the time interval between 2019 April 2 00:00:00UT and 2019 April 4 00:00:00UT. On the right and top side of the panel we show the 1D histograms of $\gamma$ and $\log_{10}(\beta)$ respectively. (Right) The 2D histogram of $\gamma$ and $\log_{10}(\beta)$, normalized to the maximum value per $\log_{10}(\beta)$ bin. In both panels, the magenta line shows the most frequent value of $\gamma$ in each $\log_{10}(\beta)$ bin.

*4.6 Polytropic Behavior and B-field orientation*

As mentioned in Section 2, *SPC* has a Sun-pointing aperture. Hence, the analysis of the obtained 1D distribution determines the radial temperature. On the other hand, the solar wind protons often exhibit large temperature anisotropies which are organized by the magnetic field direction (e.g., Marsch 2006, Verscharen et al. 2019, and references therein). In such cases, the derived temperature depends on the magnetic field direction in respect to the instrument's field of view (e.g. Huang et al. 2020). Therefore, it is important to investigate if there is a significant correlation between the polytropic index values we derive and the magnetic field direction. We examine the 8s sub-intervals within the same time period we examine in Section 4.5. For each 8s sub-interval we calculate the mean $\cos(\theta_{Br}) = B_r/B$ and its standard deviation, where $\theta_{Br}$ is the angle between the magnetic field vector and the radial direction. In the left panel of Figure 8, we show the 2D histogram of $\gamma$ and the mean $\cos(\theta_{Br})$. On the right and top side of the 2D histogram, we show the 1D histograms of the corresponding parameters. The magenta line shows the most frequent $\gamma$ in each $\cos(\theta_{Br})$ bin. The magnetic field is mainly radial ($\theta_{Br} \sim 180°$) and the occurrence drops by about a factor of 4, as $\theta_{Br}$ deviates by ~30° from the purely radial direction. Importantly, there is no obvious correlation between $\gamma$ and $\cos(\theta_{Br})$. For all the observed *B*-field directions, $\gamma \sim 2.7$, which is representative for the entire data-set we analyze through this paper. In the right panel of Figure 8, we show the 2D histogram of $\gamma$ and the standard deviation of $\cos(\theta_{Br})$. On the left and top side of the 2D histogram we show the corresponding 1D histograms of the parameters. The standard deviation of $\cos(\theta_{Br})$ within the examined sub-intervals is practically smaller than 0.1 which corresponds to ~26° from the radial direction. The magenta line shows the mode of $\gamma$ as a function of the standard deviation of $\cos(\theta_{Br})$. The most frequent $\gamma$ is bigger than 2 within the entire range of observed standard deviations of $\cos(\theta_{Br})$ and there is no significant correlation between the two parameters. We note that standard deviations associated with random fluctuations could cause small random and systematic misestimations of the mathematical calculations of $\gamma$ values via fitting (Nicolaou et al. 2019).

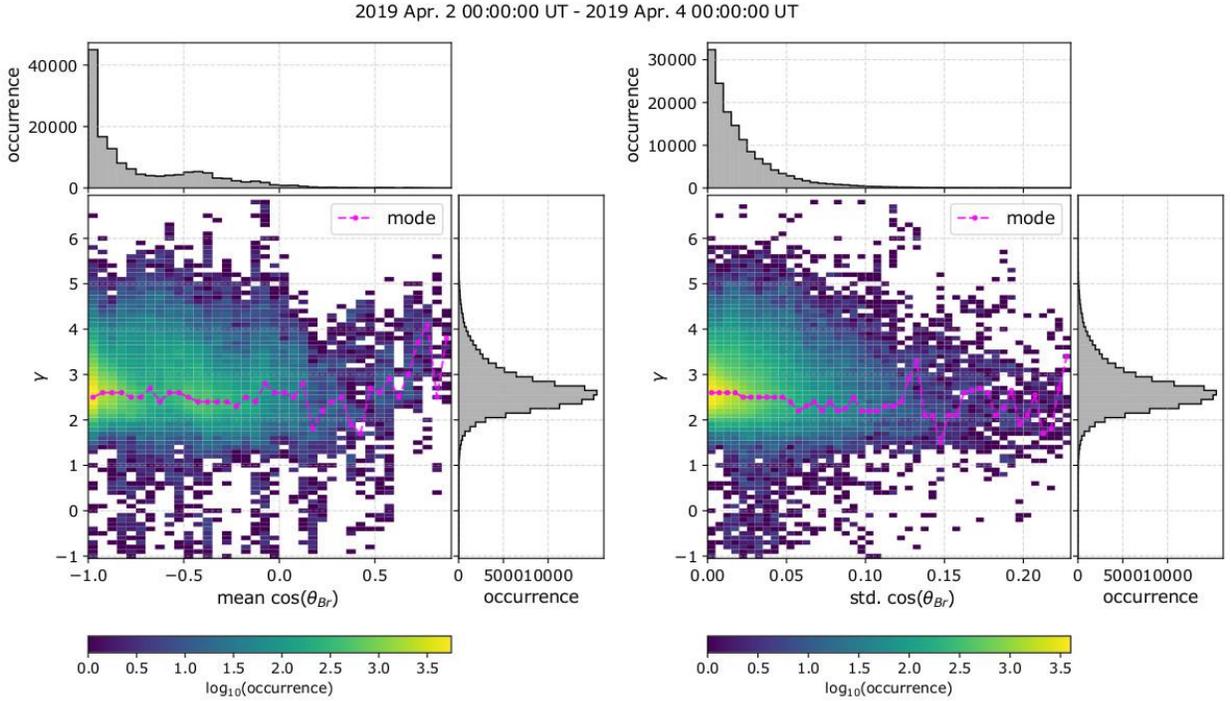

**Figure 8.** Two-dimensional histograms of (left) $\gamma$ and the mean $\cos(\theta_{Br})$ and (right) $\gamma$ and the standard deviation of $\cos(\theta_{Br})$, calculated for 8s sub-intervals between 2019 April 2 00:00:00 UT and 2019 April 4 00:00:00 UT. On the top of each 2D histogram we show the 1D histograms of the mean $\cos(\theta_{Br})$ and the standard deviation of $\cos(\theta_{Br})$ respectively, while on the right side we show the histogram of $\gamma$.

## 5. Discussion

### *5.1 Large-scale variations*

We investigate large scale variations of solar wind protons expanding in the heliosphere, as observed by *SPC* in three time intervals, between 0.17au and 1 au (Figure 1). Within these three intervals, both the plasma density and temperature vary by two orders of magnitude. The 2D histograms in Figure 3, indicate that the large-scale variations of *n* and *T*, follow a polytropic model with $\gamma \sim 5/3$, especially in the low density regime $n < 100$ cm$^{-3}$. The specific $\gamma$ corresponds to the near-adiabatic expansion model for plasma protons with three degrees of freedom ($f = 3$). In the higher density regime ($n > 100$ cm$^{-3}$) the slope of $\log_{10}(T)$ vs. $\log_{10}(n)$ seems to decrease, which indicates a possible heating of the protons, or an increase of the effective degrees of freedom. In a similar approach, Totten et al. 1995 analyze *Helios* 1 observations and determine the polytropic relation of solar wind protons from the relationships describing the plasma density and temperature as functions of the heliocentric distance between 0.3 au and 1 au. They calculate an average $\gamma \sim 1.46$, which is below the adiabatic value for $f = 3$, indicating a possible heating mechanisms acting on the protons as they expand in the heliosphere. The authors also show that the value of $\gamma$ does not depend on the solar wind speed, which is often used as a criterion to distinct between different origins of solar wind streams.

Hellinger et al. 2011 use Helios observations to study the plasma density, parallel and perpendicular temperature, speed, and magnetic field of fast solar wind protons, as functions of the radial distance within 0.3 and 1 au. The difference between their results and the predictions by the double adiabatic approach (e.g. Chew et al. 1956) makes the authors argue for a mechanism, possibly of a kinetic nature, which is cooling the plasma in the direction parallel to the magnetic field and heating it in the direction perpendicular to the magnetic field. The authors discuss a possible deceleration of secondary beams in respect to the core of the proton velocity distributions as explained in Marsch et al. 1982. Stansby et al. 2019 verify the corresponding energy transfer from the parallel to the perpendicular direction and show that alpha particles within the same heliocentric distance range exhibit a very similar behavior. The recent study of Huang et al. 2020, argues that *PSP* observes even more intense perpendicular heating and parallel cooling of the fast solar wind protons than those observed by Helios.

Here, we do not attempt to quantify a single relation between the plasma density and temperature that characterizes the entire data-set within each of the three intervals we select. Our purpose is to demonstrate the possible existence of a large-scale polytropic process characterizing the large-scale expansion of the solar wind protons in the heliosphere. Importantly, our illustrations show the existence of smaller-scale structures that deviate from the large-scale behavior.

*5.2 Profile of the plasma expansion*

The plasma density expansion within interval 1 exhibit features that clearly deviate from the $n \propto r^{-2}$ model for radial expansion with constant speed (Figure 4). A change in the radial profile exponent could be associated with a non-radial expansion geometry, or dynamic processes that decelerate or accelerate the plasma. However, it is important to note that our diagram in Figure 4 includes several streamlines crossing the spacecraft during the observations. Apparent deviations from the expected model may be due to the non-uniform streamline mixing.

We also compare the observed temperature profile with several expansion models. The models with $1.5 < \gamma < 5/3$ capture the overall trend of the observations. The large-scale profile is consistent with a nearly adiabatic or sub-adiabatic expansion for plasma protons with three degrees of freedom. However, we distinct three regions with much steeper slopes, better described by $\gamma = 2.7$, which is the typical value for the analysis of short time-scale sub-intervals. The three distinct regions we identify in Figure 4 are spread within the entire heliocentric distance range covered in interval 1.

As we show in Section 4.6 and discuss in Section 5.4 the temperature here represents the temperature tensor element along the aperture direction (radial), which is the typical magnetic field direction during the encounters.

*5.3 Short time-scale fluctuations*

In the 2D histograms of Figure 3 and the radial expansion profiles in Figure 4, we distinct sub-intervals that deviate from $\gamma \sim 5/3$ and are better described with larger $\gamma$. The statistical analysis of short time sub-intervals verifies that short time-scale fluctuations are frequently described by $\gamma \sim 2.7$ (see Figure 5). The $\gamma$ we derive is significantly larger than the values derived by previous analyses of short-time intervals (selected streamlines) at larger heliocentric distances. Newbury et al. 1997, analyze solar wind protons within stream interaction regions at ~ 0.74 au, observed by *Pioneer Venus Orbiter*. The authors analyze 73 stream interactions over time periods from 16 to 24 hours, and find intervals with $\gamma \sim 5/3$ and a few cases with $\gamma \sim 2$. The authors argue for a possible mechanism that occasionally acts on adiabatic plasma protons restricting their degrees of freedom, resulting to the few occasions with $\gamma \sim 2$. If we consider an adiabatic process with $\gamma \sim 2$, then the process is restricted to $f = \frac{2}{\gamma - 1} \sim 2$ degrees of freedom (see Section 1).

The analysis of solar wind proton streamlines at 1 au calculates a mean value of $\gamma \sim 1.8$ (e.g., Nicolaou et al. 2014b; Livadiotis 2018a,b; Nicolaou & Livadiotis 2019). These studies use proton measurements obtained with ~1min and ~1.5min resolution, and calculate $\gamma$ within sub-intervals of ~ 8 minutes. Here, although we analyze sub-intervals of similar length, we determine a significantly larger $\gamma$. We note however, that the majority of the analyzed sub-intervals in this study are obtained at $r < 0.6$ au (see Figure 1). Therefore, our results combined with the findings of the previous studies, imply that the mechanisms that trigger fluctuations with $\gamma \sim 2.7$ are possibly less effective as the plasma propagates in the heliosphere. It is also possible that at larger heliocentric distance, different streamlines with $\gamma \sim 2.7$ get well-mixed resulting to bigger streamlines with $\gamma \sim 5/3$.

In order to quantify the possible heating / cooling of the plasma, we express $\gamma$ in terms of the particle effective degrees of freedom $f = 2(c_p/c_v - 1)^{-1}$ and the ratio of the energy supplied to the system as heat to the energy supplied as work $\delta q / \delta w$ such as

$$\gamma = \frac{2}{f}\left(1 - \frac{\delta q}{\delta w}\right) + 1. \qquad (3)$$

In Figure 10, we show $\delta q/\delta w$ as a function of $f$ for different $\gamma$ values determined for solar wind protons in different studies. The red regions indicate set of values consistent with mechanisms supplying heat to the plasma ($\delta q / \delta w > 0$), while the blue regions correspond to values implying mechanisms retaining heat from the plasma ($\delta q / \delta w < 0$). Therefore, for a given $\gamma$, we can determine the range of possible effective degrees of freedom $f$ that are associated with heating and/or cooling mechanisms According to the diagrams, $\gamma \sim 2.7$ characterizes an adiabatic plasma only if $f \sim 1$. Our results are almost consistent with the kinetic description of plasma ions interacting with slow waves, where ions behave as if they are a one-dimensional adiabatic fluid with temperature variations confined along the magnetic field (e.g., Gary 1993, Verscharen et al. 2017). On the other hand, in a case with $f > 1$, our study implies the existence of a mechanism(s) cooling the plasma protons, at least in the observed direction of phase space density.

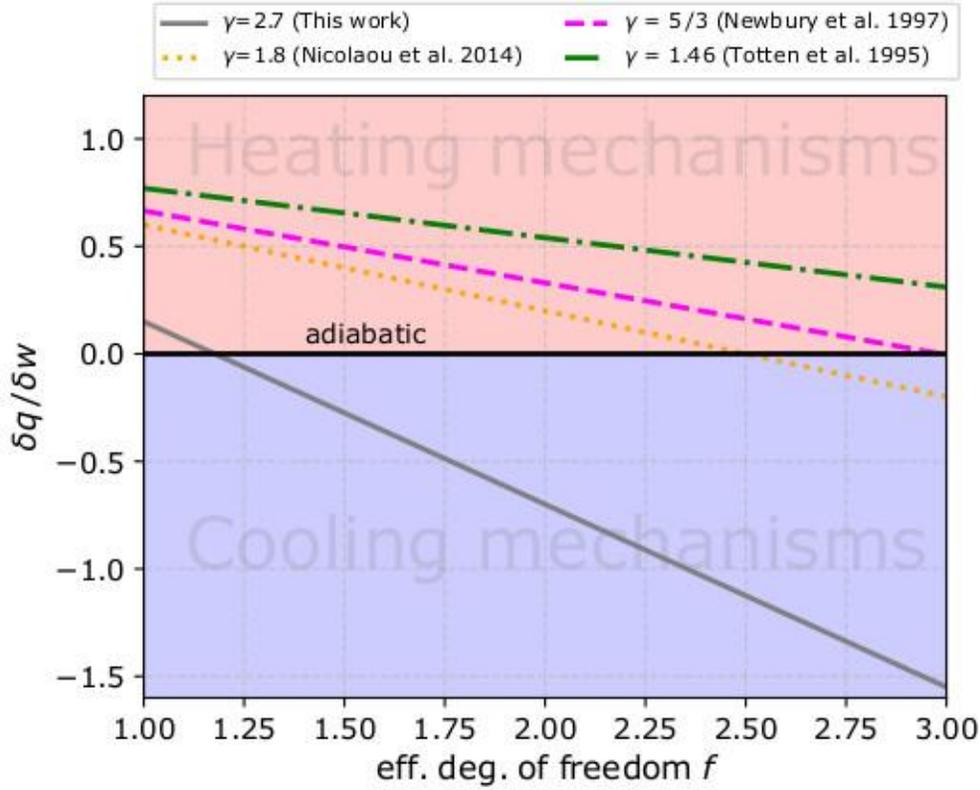

**Figure 9.** Diagrams of $\delta q / \delta w$ as a function of the effective degrees of freedom $f = 2(c_p/c_v -1)^{-1}$ for different $\gamma$ values. We explicitly show $\gamma = 2.7$ which is the typical value we derive in this work, $\gamma = 1.8$ which is the average proton polytropic index at 1 au according to Nicolaou et al. 2014, $\gamma = 5/3$ which characterizes the streams analyzed by Newbury et al. 1997, and $\gamma = 1.46$, determined by Totten et al. 1995 for solar wind protons between 0.3 and 1 au.

*5.4 Polytropic behavior and plasma properties*

We show that the polytropic index does not depend on the plasma speed. The polytropic index was found to be independent on speed by previous studies as well (Totten et al. 1995; Livadiotis 2018b; Nicolaou & Livadiotis 2019), enhancing the argument that plasma of different solar origins, do not exhibit different polytropic behavior.

On the other hand, we analyze a short-time interval in which $\gamma$ reduces with increasing plasma beta. According to our discussion in 5.2 and the diagrams in Figure 9, the observed negative correlation between $\gamma$ and $\beta$ could be due to an increase of $f$ or/and increase in $\delta q / \delta w$ with increasing $\beta$. In other words, when strong magnetic fields dominate the particle thermal motions (low $\beta$) the thermodynamic processes are confined along the direction of the magnetic field and thus, the effective degrees of freedom are reduced, or/and there is a mechanism that absorbs effectively energy from the plasma protons.

Our examination of *γ* as a function of the magnetic field direction relative to the instrument's aperture and its fluctuations does not show any significant systematic trend, eliminating the possibility of artefacts related to the 1D analysis of the plasma temperature. Additionally, our histograms in Figure 8, show that for most of the time, the magnetic field is radial and thus, the temperature values correspond mostly to the temperature parallel to the magnetic field.

## 6. Conclusions

Our study examines the large-scale and the short-scale fluctuations of the plasma density and temperature as observed by *Parker Solar Probe* between ~0.17 au and ~0.8 au. We conclude that:

- Large-scale plasma proton variations due to solar wind expansion in the heliosphere, tend to follow a polytropic model with $1.5 < \gamma < 5/3$.

- The short time-scale fluctuations associated with turbulence compressions or other local variations, follow a polytropic relation with γ ~ 2.7, indicating a process with significant cooling or/and restriction of the particle effective degrees of freedom.

- The comparison with studies at 1 au, suggests that the mechanisms that trigger fluctuations with *γ* ~ 2.7 occur predominantly near the Sun and/or short time-scale fluctuations are possibly blended in streamlines with γ~5/3 as the plasma propagates in the heliosphere.

- The polytropic index of the short time-scale fluctuations we analyze here does not depend on the plasma speed or the direction of the magnetic field.

- We analyze a case where the polytropic index exhibits a negative correlation with the plasma beta. Future statistical studies of 3D distributions can explore further this correlation and infer whether low beta decreases the effective degrees of freedom and/or favors mechanisms that retain energy from the solar wind protons in the inner heliosphere.

- Future coherent studies of the full temperature tensor of all the plasma species (protons, electrons, and heavy ions), the detailed characterization of their 3D distributions and the magnetic field, will extend our knowledge on the plasma heating mechanisms. We highlight the importance of future 3D analyses by SPAN instrument on *PSP* and the Solar Wind Analyser instrument suite (*SWA*, Owen et al. 2019) on board Solar Orbiter.

Acknowledgments

G.N., R.T.W, and D.V. are supported by the STFC Consolidated Grant to UCL/MSSL, ST/S000240/1. G.L. is supported by NASA's project 80NSSC19K0079. D.V. is supported by STFC Ernest Rutherford Fellowship ST/P003826/1. This research made use of HelioPy, a community-developed Python package for space physics (Stansby et al. 2020, http://doi.org/10.5281/zenodo.3739114). We use the publicly available SWEAP and FIELDS data found at https://spdf.gsfc.nasa.gov/pub/data/psp. We acknowledge the NASA Parker Solar Probe Mission, SWEAP team led by Justin Kasper and Fields team led by Stuart Bale for use of data.# References